\documentclass{Interspeech2024}
\usepackage{multirow}
\usepackage{algorithm}
\usepackage{algorithmic}

\interspeechcameraready

\title{Simul-Whisper: Attention-Guided Streaming Whisper with Truncation Detection}

\name{Haoyu Wang$^{1\text{†}}$, Guoqiang Hu$^{2\text{†}}$, Guodong Lin$^{1\text{†}}$, Wei-Qiang Zhang$^{1*}$, Jian Li$^3$}

\address{
  $^1$Department of Electronic Engineering, Tsinghua University, Beijing 100084, China \\
  $^2$International School, Jinan University, Guangzhou, 511443, China \\
  $^3$Beijing Sinovoice Technology Co.,Ltd, China}
\email{\thanks{\text{\dag } Equal contribution}\thanks{* Corresponding author}\thanks{This work was supported by the National Natural Science Foundation of China under Grant No. 62276153.}w-hy21@mails.tsinghua.edu.cn, %violetpark6567@gmail.com,
wq-zhang@tsinghua.edu.cn}


\begin{document}

\maketitle

\begin{abstract}
As a robust and large-scale multilingual speech recognition model, Whisper has demonstrated impressive results in many low-resource and out-of-distribution scenarios. However, its encoder-decoder structure hinders its application to streaming speech recognition. In this paper, we introduce \textbf{Simul-Whisper}, which uses the time alignment embedded in Whisper's cross-attention to guide auto-regressive decoding and achieve chunk-based streaming ASR without any fine-tuning of the pre-trained model. Furthermore, we observe the negative effect of the truncated words at the chunk boundaries on the decoding results and propose an integrate-and-fire-based truncation detection model to address this issue.  Experiments on multiple languages and Whisper architectures show that \textbf{Simul-Whisper} achieves an average absolute word error rate degradation of only 1.46\% at a chunk size of 1 second, which significantly outperforms the current state-of-the-art baseline.
\end{abstract}

\section{Introduction}

Pre-training on large-scale datasets has brought significant improvements in automatic speech recognition (ASR) \cite{mohamed_self-supervised_2022, zhao_improving_2022}. Recently, Whisper \cite{radford2023robust}, a weakly supervised encoder-decoder model trained on a large dataset of 680,000 hours, was shown to have remarkably robust performance across different languages and complex open environments \cite{radhakrishnan-etal-2023-Whispering, gong_Whisper-at_2023, ma2024investigating, rathod_noise_2023}. On the other hand, Whisper is not pre-trained to perform streaming ASR, which sets a barrier to its application in scenarios such as conference transcription, simultaneous translation, and live streaming. This makes using Whisper for streaming ASR an attractive proposition. 

Compared to offline ASR, streaming ASR is more challenging due to the lack of full context during inference. Transformer-based streaming ASR models typically use time-restricted \cite{zhang2020TT}, chunk-wise \cite{tian2020st}, or memory-based \cite{shi_emformer_2020} attention. However, modifying the pre-trained parameters to use such streaming attention mechanisms requires large computational resources \cite{fu_distillw2v2_2023}. Besides, compared to encoder-only pre-trained models such as HuBERT \cite{hsu2021hubert} or WavLM \cite{chen2022wavlm}, Whisper models face additional challenges when used for streaming ASR due to their encoder-decoder structures. In an encoder-decoder structure, the encoder converts the input audio into latent representations all at once, after which the decoder begins iterative auto-regressive decoding until it encounters a special 
 `` $\langle \text{eos} \rangle$'' token in the output sequence. In streaming ASR, the input to the model is often truncated into short segments of fixed duration. This random truncation can lead to unreliable outputs at the end of the transcription, and can even cause attention to fail, resulting in meaningless repetitive outputs. These errors cannot be avoided by stopping decoding before the boundary, since the start and end timestamps corresponding to the output tokens cannot be found due to the non-monotonic nature of the encoder-decoder cross-attention.
 
\begin{figure}[t]
  \centering
  \includegraphics[width=0.9\linewidth]{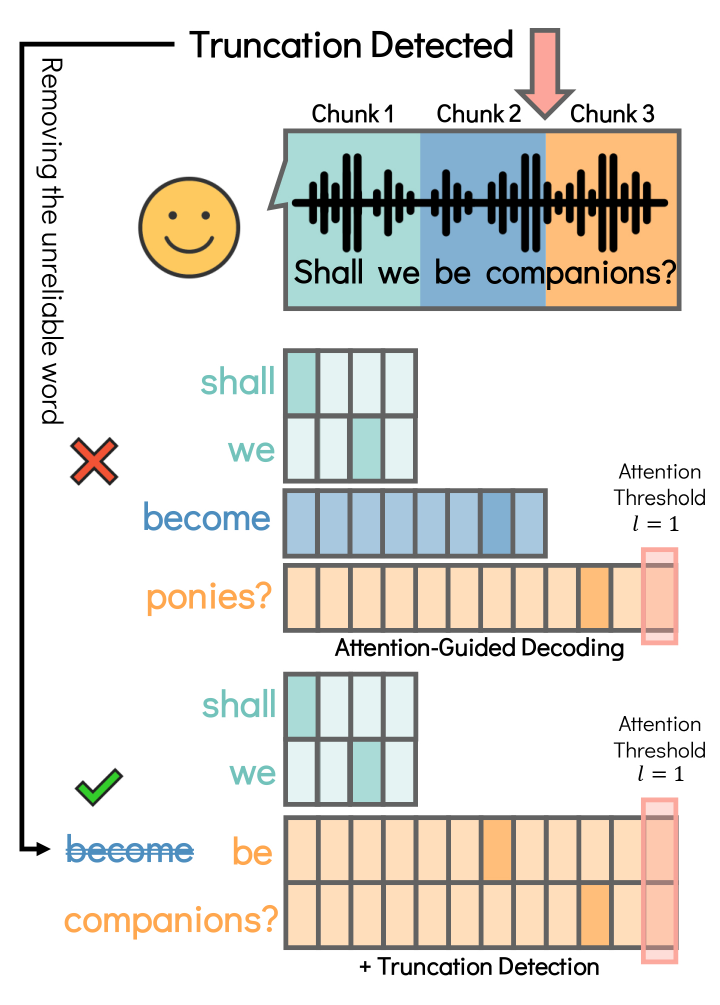}
  \caption{An overview of our method. Different colours indicate different audio chunks and their corresponding transcriptions. Darker blocks in the cross-attention matrix indicate the audio frames most attended to by the current token. Upper part: Decoding is stopped when the most attended audio frame appears at the chunk boundary. Lower part: The unreliable last word is deleted from the transcription when truncation is detected and the model waits until the next chunk is received.}
  \label{fig:method}
\end{figure}

To address these issues, Macháček et al. propose to apply the Local Agreement \cite{machavcek2023turning} policy to the Whisper models. They use a chunk-based inference approach, where Whisper performs non-streaming decoding each time a chunk of audio is received, and the final transcription is determined by the longest common prefix of the results from previous chunks. This method avoids modifying the parameters of Whisper, while mitigating unreliable transcriptions at chunk boundaries. However, since the model only generates transcriptions when the longest common prefix grows, the latency is usually unpredictable. Additionally, repeatedly performing non-streaming decoding is also computationally intensive.

Streaming inference of encoder-decoder models is also an important topic in speech translation \cite{berard2016listen, weiss2017sequence, ma2019monotonic}. Recently, some studies have proposed the direct use of non-streaming models for streaming translation \cite{papi-etal-2022-simultaneous, papi2023alignatt, ma2018stacl, polak-etal-2022-cuni}, achieving close or even better performance than models requiring streaming-aware training. These works focus on the nearly monotonic cross-attention within appropriately trained encoder-decoder models, enabling control over the time to stop decoding. For example, EDA\begin{scriptsize}TT\end{scriptsize} \cite{papi_attention_2023} stops decoding when the attention on the last few audio frames reaches a certain threshold, while AlignAtt \cite{papi2023alignatt} focuses on the position of maximum attention and stops decoding when it is close enough to the end of the audio. Compared to the Local Agreement method, this type of method requires less computation and has better control over decoding latency. Since Whisper's cross-attention also shows good temporal alignment, we believe that this type of fine-tuning-free method can also be applied to Whisper's streaming inference.

However, a problem with these methods is that cross-attention primarily provides information from the decoder. As shown in Figure \ref{fig:method}, truncated words at chunk boundaries can lead to wrong transcriptions, and relying only on decoder information may not be sufficient to discriminate such cases. To exclude these unreliable transcriptions, information from the encoder is crucial. Recently, Dong et al. \cite{dong2022learning} and Zhang et al. \cite{zhang2022information} propose to train an integrate-and-fire (IF) model \cite{abbott1999lapicque, dong2020cif, yi2021efficiently} to detect the word boundaries. Similar methods can be applied to word truncation detection, or in other words, if a word boundary is not detected at the end of a chunk, it is equivalent to a word truncation.

In this paper, we propose Simul-Whisper, a streaming inference strategy for Whisper that does not require fine-tuning of the pre-trained model\footnote{Available at \url{https://github.com/backspacetg/simul\_whisper}}. By integrating the encoder and decoder information from both the cross-attention and the truncation detection module, we can track the decoding process, stop decoding at appropriate times, and discard unreliable transcriptions. Our experiments on multiple languages and Whisper architectures demonstrate that the proposed method achieves streaming inference with a chunk size of 1 second with a minimum absolute performance degradation of 0.77\%, which significantly outperforms the state-of-the-art Local Agreement baseline.

\section{Methods}
In this section, we introduce our method for streaming inference using non-streaming Whisper without any fine-tuning. First, we utilise Whisper's cross-attention mechanism to obtain a rough alignment and stop decoding at the appropriate time. Second, to further eliminate unreliable transcriptions at chunk boundaries, we introduce a Truncation Detection Module (TDM) based on the Integrate-and-Fire (IF) mechanism. 

\subsection{Attention-Guided Decoding Policy}

Whisper has various model architectures, all of which consist of a 2-layer CNN, along with a multi-layer Transformer encoder and decoder. The encoder-decoder attention is calculated by

\begin{align}
    Q &= X_tW^i_Q \\
    K&=X_aW^i_K\\
    S^i&=\text{softmax}(\frac{QK^{T}}{\sqrt{d_K}})\text{, }
\end{align}where $X_a \in \mathbb{R}^{N_a \times D_a}$ is the encoded audio, $X_t$ in the previous outputs of the decoder. $N_a$ and $D_a$ are the length and hidden dimension of the encoded audio, respectively. Similarly, $N_t$ and $D_t$ are those for the the decoder outputs. $S^i$ in $\mathbb{R}^{N_t \times N_a}$ is the output of the $i$th attention head of the multi-head cross-attention module. 

During the large-scale weakly supervised training of Whisper, some attention heads in the cross-attention modules exhibit a favourable temporal alignment. To be specific, for the $t$th row $s^i_t$ of the attention matrix, the region where token $t$ shows significant attention to $X_a$ often substantially overlaps with the corresponding audio. These attention heads are manually selected and called alignment heads, and the timestamps generated by Whisper are actually obtained by applying Dynamic Time Warping (DTW) to them. Following the open source code of Whisper\footnote{https://github.com/openai/whisper}, we perform further post-processing on the output of the alignment heads. The resulting alignment matrix can be represented as:

\vspace{-1em}
\begin{align}
S &= f_m(\sum_{i \in H}s^i) \text{,}
\end{align} where $f_m$ is a median filter with a window width of 7, $H$ is the set of the alignment heads. To guide the decoding process, we need to find out from the alignment head the position where the current token focuses on in the audio. Inspired by Papi et al. we use the position of the maximum attention as a reference. To be specific, the decoding process will be terminated when

\vspace{-1em}
\begin{align}
N_a - \text{argmax}(S_t) < l
\end{align} to avoid incorrect or repetitive tokens, where $l$ is a predefined threshold. Compared to DTW, this approach treats each token as an independent unit, thereby reducing the cumulative error in the auto-regressive decoding progress. Besides, selecting the maximum value can also reduce the noise caused by the randomness of the model. It is therefore more suitable for tracking the streaming inference process where training and testing are significantly mismatched. 

\subsection{IF-Based Truncation Detection Module} 
The fixed-length audio chunks provided to the model may contain incomplete speech units, which leads to unreliable transcriptions. Moreover, these errors accumulate and affect the whole sentence. To address these issues, we design a IF-based truncation detection module. If a truncation is not detected during decoding, the resulting token is preserved. Otherwise, the unreliable token is removed and generated when the complete word is received in the next chunk.

The IF neuron continuously receives and integrates signals. When the accumulated signal exceeds a certain threshold, the neuron fires, or in other words, generates an output and resets the accumulated value. In our TDM, the signal is generated by mapping the output of Whisper's encoder through a simple linear layer and a sigmoid function. Our training objective is to ensure that the number of firings of the IF neurons matches the number of words in the audio. During inference, a truncation is detected if the IF neuron does not fire at the end of the audio chunk. Given an IF threshold $f$, signal sequence $a \in \mathbb{R}^{N_a}$, accumulated signal $I$, the TDM will record the last position $p$ where IF fires to detect truncation according to the operations below:

\begin{algorithm}
\caption{truncation detection module}
\begin{algorithmic}[1]
    \STATE \textbf{Input:} $\alpha$
    \STATE Initialize $I \gets 0$, $p \gets 0$
    \FOR{$n = 0$ \TO $N_a-1$}
        \STATE Compute $I \gets I + \alpha_{n}$
        \STATE Determine $\text{fire}$ where $I\geq f$
        \STATE Update $I$: 
        \STATE \quad $I\gets \begin{cases} I - f, & \text{if fire}\\ I, & \text{otherwise} \end{cases}$
        \STATE Update $p \gets n$
    \ENDFOR
    \STATE Determine $\text{truncation}$ where $p \leq N_a-1$
\end{algorithmic}
\end{algorithm}

Since Whisper pads the input to 30 seconds during inference, the last frame of the encoder features is actually the transition point from speech content to silent padding, where the IF neuron always fires. Therefore, to ensure the correctness of the IF results, we discard the last frame of the encoder features.

\section{Experimental Settings}

\subsection{Data}
The proposed method is evaluated on the LibriSpeech dataset \cite{panayotov2015librispeech} and the Multilingual Librispeech (MLS) dataset \cite{pratap2020mls}, which consists of seven languages, including Dutch (nl), French (fr), Polish (pl), German (de), Italian (it), Portuguese (pt), and Spanish (es). For the Librispeech dataset, the subsets test-clean, dev-clean, test-other and dev-other are evaluated, and for MLS we only use the test subsets. 

\subsection{Truncation Detection Training Setup}
The root mean square error (RMSE) is used as the training objective of the TDM, and our training dataset is the 100-hour Librispeech train-clean subset. The Adam optimizer is used together with warm-up to prevent unstable behaviour in the early stages of training. Our TDM is trained for 10 epochs with a batch size of 4500 MFCC frames, and the first 3 epochs are used for warm-up, where the learning rate increases linearly from 0 to $1 \times 10^{-6}$. Training is performed on a single NVIDIA GeForce RTX 3090 GPU with 24GB of memory. 

\subsection{Inference and Evaluation}
The generated transcriptions and ground truth labels are normalized using Whisper's open source text normalizer, and the Python library editdistance \cite{yujian2007normalized} is used to evaluate the word error rates. We use a threshold of $l=12$ frames (240 ms) for attention-guided decoding, and the fire threshold for TDM is 0.999. For the baseline, we reproduce the Local Agreement policy based on its open source code\footnote{https://github.com/ufal/whisper\_streaming} and use the experimental setup described in the original paper, which computes the longest common prefix between $n=2$ consecutive source chunks. We use the Differentiable Average Lagging (DAL) \cite{arivazhagan2019monotonic} to evaluate the latency. All inferences are performed on a single NVIDIA GeForce RTX 3090 GPU with 24GB of memory.

\section{Results}

\begin{figure}[t]
  \centering
  \includegraphics[width=0.9\linewidth]{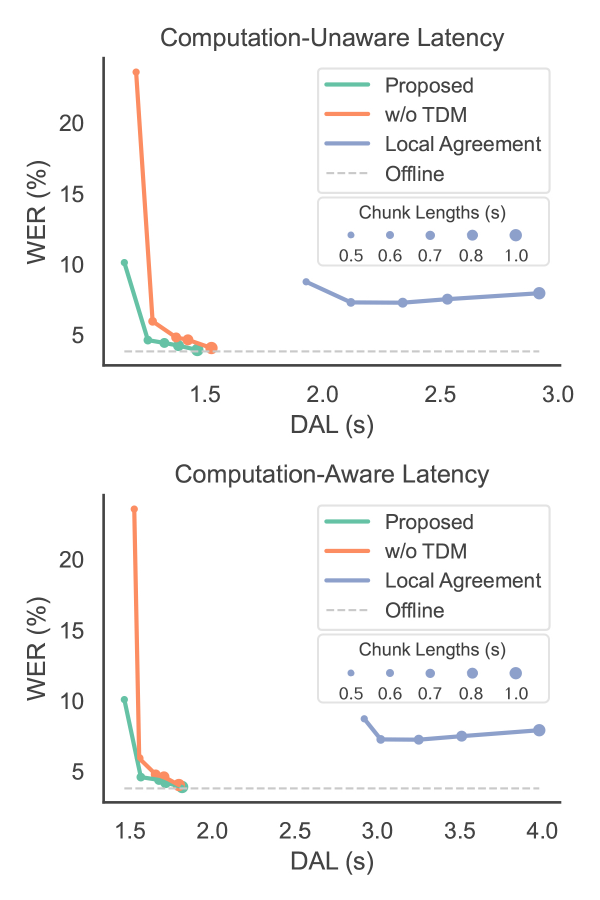}
  \vspace{-1em}
  \caption{The WERs at different DALs for Whisper Large-v2, where computation-aware latency takes into account processing time, while computation-unaware latency doesn't. Chunk lengths vary from 0.5 to 1.0 seconds. For the same latency, the proposed method exhibits a significantly lower WER compared to the baseline, and the addition of IF further improves performance.}
  \vspace{-1em}
  \label{fig:delay}
\end{figure}

\subsection{The Word Error Rates of Simul-Whisper}

Table \ref{tab:main} shows the WERs for offline decoding and streaming decoding policies on the Librispeech and MLS datasets, where the chunk length for streaming decoding is 1 second.  In these experiments, the proposed method achieves streaming decoding with a minimum absolute performance degradation of 0.09\%. Generally, the Medium and Large architectures exhibit less performance degradation than the Base and Small architectures. Moreover, the streaming WERs on Librispeech test-clean and dev-clean datasets are usually lower than those in MLS. This observation suggests that the performance degradation in streaming decoding might be related to the model's initial performance. During streaming inference, the input audio is often short and randomly truncated, necessitating the original offline model to be robust.

\begin{table*}[htb!]
    \caption{The WERs (\%) of streaming decoding policies on the Librispeech and MLS datasets, with a chunk length of 1 second. $\bar{\Delta}$ is the average performance degradation. The proposed method achieves a minimum performance degradation of 0.09\%, significantly outperforming the Local Agreement baseline. Using TDM exhibits positive effects on almost all datasets and model architectures.}
    \vspace{-1em}
    \resizebox{\textwidth}{!}{
    \centering
    \begin{tabular}{lccccccccccccc}
        \toprule
         \multirow{2}*{\textbf{Architectures}} & \multirow{2}*{\textbf{Methods}} & \multicolumn{4}{c}{\textbf{Librispeech}}  &  \multicolumn{7}{c}{\textbf{MLS}} & \multirow{2}{*}{\textbf{$\bar{\Delta}$ (↓)}} \\ 
        \cmidrule{3-13}
         &  & test-clean & dev-clean & test-other & dev-other & nl & fr & pl & de & it & pt & es & \\ 
        \midrule 
        \multirow{4}*{\textbf{Base}} & Non-streaming & 5.89  & 6.16  & 12.91  & 12.21  & 30.84  & 25.10  & 25.20  & 19.90  & 32.80  & 23.40  & 14.40 & - \\ 
        & Local Agreement & 12.73  & 11.54  & 19.40  & 19.26  & 44.23  & 38.62  & 37.25  & 41.01  & 45.52  & 33.94  & 24.80 & 10.86 \\ 
        & Proposed & 8.07  & 8.55  & 18.58  & 19.16  & 36.47  & 37.96  & 33.46  & 25.84  & 44.52  & 39.56  & 21.08 & \textbf{7.68}  \\ 
        & w/o TDM & 9.35  & 9.08  & 19.56  & 19.33  & 37.39  & 37.68  & 33.46  & 26.57  & 43.46  & 37.76  & 20.31 & 7.74 \\ 
        \midrule
        \multirow{4}*{\textbf{Small}} & Non-streaming & 4.18  & 4.41  & 8.89  & 8.40  & 18.17  & 13.74  & 12.10  & 11.51  & 26.55  & 13.84  & 8.91 & - \\ 
         & Local Agreement & 9.29  & 9.48  & 14.20  & 13.83  & 27.69  & 22.17  & 21.39  & 21.10  & 27.00  & 23.68  & 17.12 & 6.93 \\ 
         & Proposed & 5.15  & 5.44  & 11.39  & 11.18  & 25.87  & 18.05  & 16.83  & 17.36  & 27.21  & 16.73  & 10.96 & \textbf{3.23} \\ 
         & w/o TDM & 6.23  & 6.14  & 13.13  & 12.65  & 26.60  & 19.54  & 18.79  & 19.04  & 29.64  & 19.63  & 12.73 & 4.86 \\ 
         \midrule
        \multirow{4}*{\textbf{Medium}} & Non-streaming & 3.80  & 3.78  & 8.50  & 6.84  & 12.40 & 9.43 & 6.78 & 8.24 & 15.51 & 8.89 & 5.93 & - \\ 
        & Local Agreement & 14.61  & 13.46  & 16.24  & 15.77  & 14.76  & 20.51  & 42.11  & 18.00  & 21.57  & 17.55  & 12.08 & 10.59 \\ 
        & Proposed & 3.90 & 4.02 & 8.12 & 7.70 & 15.05 & 10.30 & 11.40 & 10.20 & 18.13 & 11.05 & 7.66 & \textbf{1.46} \\ 
        & w/o TDM & 4.36 & 4.26 & 9.23 & 8.74 & 18.37  & 12.15  & 10.43  & 11.35  & 21.03  & 13.40  & 7.90 & 2.83 \\ 
        \midrule
        \multirow{4}*{\textbf{Large-v2}} & Non-streaming & 3.79  & 3.76  & 7.38  & 6.70  & 9.68  & 6.82  & 5.49  & 6.26  & 13.14  & 6.63  & 4.78 & - \\ 
         & Local Agreement & 7.91  & 6.91  & 11.47  & 11.41  & 18.89  & 15.13  & 13.44  & 14.73  & 20.73  & 16.35  & 13.39 & 6.90 \\ 
         & Proposed & 3.89 & 3.85 & 8.89 & 8.14 & 12.86  & 9.66  & 7.90  & 8.94  & 17.58  & 12.30  & 6.21 & \textbf{2.34} \\
         & w/o TDM & 4.03 & 3.96 & 9.13 & 8.56 & 15.05  & 11.19  & 9.80  & 13.70  & 18.13  & 12.92  & 8.08 & 3.65 \\  
         \bottomrule
    \end{tabular}
    }
    \label{tab:main}
    \vspace{-1em}
\end{table*}

We also compare the average performance degradation $\bar{\Delta}$ on all the datasets above. Our proposed method exhibits significantly lower $\bar{\Delta}$ compared to the baseline. We observed that the transcription of the Local Agreement policy often contains insertion or deletion errors at the ends of sentences, which may be related to its context management strategy. Local Agreement retains a variable-length audio context, and when Whisper outputs a delimiter (such as a period), the buffer is truncated from the delimiter based on Whisper's timestamp. Hence, errors may occur if the timestamp is not precise enough. On the other hand, the attention-guided policy does not require precise timestamps. We retain the previous audio segment and its transcription result and insert them into a queue. When the length of the retained context exceeds a fixed threshold, the audio and transcription at the top of the queue are removed. Although the remaining transcriptions slightly precede the current audio, this additional context has less negative effect than incorrectly truncated audio. Furthermore, we use the context as conditional input to the decoder, which can better control the decoding than using it as a prompt in Local Agreement.

Besides, the proposed IF-based truncation detection module has a positive effect on accuracy across almost all model architectures and languages, indicating the generalization ability of our method. For Whisper Medium and Large, Simul-Whisper has comparable performance on Librispeech test-clean and dev-clean datasets, suggesting that the performance degradation in these datasets may be mainly caused by truncated words without TDM.

\subsection{The Latency of Simul-Whisper}

We use Differentiable Average Lagging (DAL) to estimate the latency of streaming policies. DAL is the latency relative to the ideal streaming system averaged over all tokens. Assuming the input audio length is $N_a$ and the number of output tokens is $N_t$, the ideal streaming policy generates a token every $d=N_a/N_t$ seconds. Assuming that token $t$ is generated at time $g(t)$, DAL is calculated as follows:

\begin{align}
    g\prime_d(t) = \begin{cases}
        g(t)&\text{ }t=1 \\
        \text{max}(g(t), g\prime_d(t-1)+d)&\text{ }t>1
    \end{cases}
\end{align}
\vspace{-1em}
\begin{align}
    \text{DAL}=\frac{1}{N_t}\sum^{N_t}_{t=1}g\prime_d(t)-(t-1)d\text{.}
\end{align}

By adjusting $g(t)$ to $g\prime(t)$, meaningless negative values can be avoided. We evaluate the computation-unaware and computation-aware latency. For the computation-unaware latency, we ignore the computation time and assume that the output is available immediately after a chunk is received, while for the computation-aware latency, the hardware processing time of the model is included.

Figure \ref{fig:delay} shows the latency of the Simul-Whisper and baseline models, with chunk lengths varying from 0.5 to 1.0 seconds. It can be seen that for the same latency, the WER of the proposed model is significantly lower than that of the Local Agreement baseline. Local Agreement has a higher latency because it only generates tokens as the longest common prefix grows, and the latency cannot be directly controlled by the chunk length. On the other hand, the Simul-Whisper's latency is generally 1 to 2 times the chunk length. This is because most tokens are either generated after the current chunk (1-chunk latency) or delayed until the next chunk (2-chunk latency). Finally, although TDM slightly increases computation time, it achieves less performance degradation for the same computation-aware latency, demonstrating its effectiveness.

%\vspace{-0.5em}
\section{Conclusions and Discussions}

In this paper, we introduce Simul-Whisper, a streaming decoding policy for Whisper without additional fine-tuning of the pre-trained model. We utilise cross-attention to guide the decoding process and train a truncation detection module to avoid unreliable transcriptions at chunk boundaries. Experiments on multiple datasets show that the proposed method can achieve streaming ASR with an average absolute performance degradation of 1.46\% at a chunk size of 1 second.

There are still some problem within Simul-Whisper. Since we prefer a fine-tuning-free method, inputs still need to be padded to 30 seconds to meet Whisper's input requirements, which increases computation-aware latency. In future work, we will explore methods such as self-distillation to address this issue.

\clearpage

\bibliographystyle{IEEEtran}
\bibliography{bib_mod}

\end{document}